\documentclass[aps,preprint,nofootinbib,floatfix]{revtex4-2}
\usepackage[utf8]{inputenc}

\usepackage[title]{appendix}
\usepackage{hyperref}
\usepackage{graphicx}
\usepackage{amssymb}
\usepackage{amsmath}
\usepackage{subfigure}
\usepackage{bm}
\usepackage{booktabs}
\usepackage{xcolor}

\begin{document}
\title{Phenomenological study of heavy neutral gauge boson in the left-right symmetric model at future muon collider}
\author{Zongyang Lu$^{a,b}$}
\author{Jianing Qin$^{a}$}
\author{Honglei Li$^{a}$}
\email{sps\_lihl@ujn.edu.cn}
\author{Zhi-Long Han$^{a}$}
\email{sps\_hanzl@ujn.edu.cn}
\author{Fei Huang$^{a}$}
\email{sps\_huangf@ujn.edu.cn}
\author{Chun-Yuan Li$^{c}$}
\author{Xing-Hua Yang$^{c}$}
\author{Zhong-Juan Yang$^{a}$}

\affiliation{$^a$School of Physics and Technology, University of Jinan, 250022, Jinan, China}
\affiliation{$^b$Key Laboratory of Particle Physics and Particle Irradiation (MOE),
Institute of Frontier and Interdisciplinary Science,Shandong University, Qingdao, Shandong 266237, China}
\affiliation{$^c$School of Physics and Optoelectronic Engineering, Shandong University of Technology,  255000, Zibo, China}
\begin{abstract}
The exotic neutral gauge boson is a powerful candidate  for the new physics beyond the standard model.  As a promising model, the left-right symmetric model has been proposed  to explain the neutrino mass, dark matter, and matter-antimatter asymmetry, etc., in which exotic gauge bosons $Z^\prime, W^{\prime \pm}$ have been put forward as well as other new right-handed particles. We investigate the  $\mu^+ \mu^- \to q\bar{q} $ and $ \mu^+ \mu^- \to l^+ l^- $ processes involving the $Z^\prime$ boson as an intermediate particle. 
The  coupling strength, decay width and mass are the key parameters on the production and decay processes of the $Z^\prime$ boson. The results indicate that the angular distributions of final particles are sensitive to the couplings of $Z^\prime$ to the other fermions.  Asymmetries defined from the angular distributions are ideal quantities to demonstrate the discrepancy between the standard model process and the processes with $Z^\prime$ participated and they are also appropriate observables to discriminate the couplings of $Z^\prime$ to other particles. Compared with the current results at the Large Hadron Collider (LHC), the future muon collider has a great potential to explore the new  parameter space with  $Z^\prime$ boson.
\end{abstract}

\maketitle
\newpage
\quad
\section{Introduction}
In the realm of particle physics, the Standard Model (SM)~\cite{Thomson:2013zua} has garnered considerable success by providing accurate descriptions of known elementary particles and their interactions. However, with the continuous accumulation of experimental data and the deepening of research, people are increasingly realizing that there are unresolved issues in the SM. Some of these issues may be related to an extension of the gauge group  beyond the SM, involving extra neutral gauge bosons ($Z^\prime$). 
 As a category of particles that are massive and participate in the gauge interactions, the $Z^\prime$ boson has been predicted in numerous models~\cite{Amrith:2018yfb,Kim:2016plm,Frank:2019nid,Cao:2016uur}. For instance, the Grand Unified Theory (GUT)~\cite{Dimopoulos:1981zb,Beasley:2008kw,Langacker:2008yv,delAguila:2010mx,Nath:2010zj} and models introducing an additional $U(1)_X$ group \cite{Babu:1997st,Wells:2008xg,Gershtein:2008bf} are among them. Despite the fact that in various models, the $Z^\prime$ originates from distinct group structures, it is a common feature that $Z^\prime$ typically couples with Standard Model fermions~\cite{Erler:2009jh}. This linkage with Standard Model fermions provides a  foundation for investigating neutral gauge bosons. In other words, we can study the coupling of $Z^\prime$ bosons with fermions by examining the distribution of final-state fermions, the scattering cross section of the process, and other physical properties. This allows for an investigation of the characteristics of $Z^\prime$ bosons in different models and their potential behavior at colliders. By simulating collider experiments, one can obtain rich phenomenological properties of $Z^\prime$ bosons through collider experiments. This research can help us understand the differences in the properties of $Z^\prime$ bosons in various models, thereby enhancing the detection efficiency and accuracy of neutral gauge bosons at colliders.
 
To date, many collider experiment collaborations worldwide are  attempting to search for signals of neutral gauge bosons. Based on various theoretical models, different experimental groups have provided different mass constraints on the $Z^\prime$ boson~\cite{CDF:2008ieg,Dobrescu:2014fca,CMS:2018rkg,CMS:2019xai,CMS:2019emo,CMS:2019gwf,ATLAS:2022enb,CMS:2023hwl,CMS:2023byi}. Specifically, the maximum exclusion limits are less than 4000 GeV for $Z^\prime \to Zh$ and $Z^\prime \to W^+ W^-$~\cite{CMS:2021klu}, less than 4500 GeV for $Z^\prime \to l^+ l^-$~\cite{ATLAS:2018sbw}, 1800 GeV - 2400 GeV or 600 GeV - 2100 GeV for $Z^\prime \to b \bar{b}$~\cite{CMS:2022eud,ATLAS:2018tfk}, and less than 3900 GeV for $Z^\prime \to t \bar{t}$~\cite{ATLAS:2020lks}. 
 However, due to the limitations in the luminosity and collision energy of current colliders, detecting high-mass $Z^\prime$ bosons remains challenging. Currently, many new physics models with a few TeV scale neutral gauge bosons have been tested at the LHC. The prospect of detecting $Z^\prime$ neutral gauge bosons with higher masses may hinge upon the advent of future large hadron colliders or muon colliders. Fortunately, the development is that future high-energy hadron colliders and muon colliders have already entered the stage of consideration. The design collision energy of the Muon Collider (MuC) exceeds 10 TeV~\cite{Accettura:2023ked} and the design collision energy of the Future Circular Collider (FCC) reaches 100 TeV~\cite{FCC:2018evy}. This holds important significance for the detection of heavy neutral gauge bosons and the elucidation of their intrinsic properties.

In prior investigations, we conducted a comprehensive phenomenological study of models incorporating an additional $U(1)_X$ group, including an examination of various potential decay channels of the $Z^\prime$ boson~\cite{Yin:2021rlr,Li:2013ava}. However, a notable observation emerged, indicating a prevalent focus in existing research on the neutral gauge boson within the $Z^\prime$ framework, predominantly associated with an additional $U(1)$ group, be it $U(1)_X$ or $U(1)^\prime$. The existence of a right-handed neutral gauge boson within an additional $SU(2)_R$ group continues to evoke interest. Studies aimed at those exploring left-right symmetric models often emphasize the investigation of additional $W^\prime$ bosons or the introduction of extra right-handed neutrinos, potentially associated with dark matter, while overlooking the phenomenological properties of the $Z^\prime$ coupling with fermions. This paper endeavors to address this gap by conducting a detailed examination of the interaction properties between a high-mass $Z^\prime$ boson within a left-right symmetric model and SM fermions. This provides a basis for a detailed understanding of neutral current processes within left-right symmetric models.

Muons, like protons, can be accelerated within a relatively compact ring and collide at center-of-mass system (C.M.S) energies of a few TeV (e.g., 3 TeV, 6 TeV, 10 TeV, or even higher), without being constrained by synchrotron radiation. In recent years, as various issues encountered in the construction of a muon collider have been addressed, the development of the muon collider has entered the preparation phase. The International Muon Collider Collaboration (IMCC) has been actively exploring the construction of a muon collider with a C.M.S energy exceeding 10 TeV and higher luminosity~\cite{Accettura:2023ked,Ruhdorfer:2023uea}. While the motivation for constructing a muon collider may stem from the prospect of establishing a Higgs factory~\cite{Ankenbrandt:1999cta,AlAli:2021let,Han:2020pif,Forslund:2022xjq,Bartosik:2019dzq,Bartosik:2020xwr}, it is essential to acknowledge that muon colliders also offer distinct advantages in the detection of heavy resonances with greater mass. Compared with the future electron-positron collider and hadron collider, the superiority of the muon collider is evident in  collision energy and background cleanliness, underscoring the broader scientific potential of the muon collider beyond its role as a Higgs factory.

This paper is organized as follows. In Section 2, we will give a detailed introduction to the theoretical framework. In Section 3, we will conduct a comprehensive study on the scattering cross sections, angular distributions of final-state particles, as well as the asymmetry in angular distributions associated with different $Z^\prime$ decay processes.  Finally, a summary is presented.

\section{Theoretical Framework and Collider Constraints}
In order to elucidate the origins of neutrino mass, the existence of dark matter, and asymmetry of positive and negative matter in the universe, the Left-Right Symmetric Model (LRSM)~\cite{Pati:1974yy,Mohapatra:1974gc,Senjanovic:1975rk} has been proposed. This model is founded on the gauge group 
\begin{equation}
	SU(3)_c \times SU(2)_L \times SU(2)_R \times U(1)_{B-L},
\end{equation}
offering a theoretical framework to comprehensively address these fundamental cosmological phenomena. It predicts the existence of right-handed (RH) currents and heavy  gauge bosons $W^{\prime \pm}$ and  $Z^\prime$. Additionally, the model incorporates three RH neutrinos, which carry electric charge under $SU(2)_R \times U(1)_{L-R}$ and yet remain singlets under the Standard Model symmetries. The LRSM has been extensively investigated and constraints on certain particles have been provided through processes such as $p p \to W^{\prime \pm} \to tb$ or $N_R l^{\pm}(N_R \to l^{\pm} W^{\prime \mp} \to l^{\pm} q \bar{q}^\prime)$ \cite{Mitra:2016kov,Han:2012vk} and $p p \to Z^\prime \to t \bar{t}$ or $N_R N_R$ \cite{Frederix:2021zsh,CMS:2023ooo} at LHC. The kinematic changes induced by high-transverse momentum ($p_T$) initial-state radiation (ISR) or final-state radiation (FSR) are investigated in~\cite{Mattelaer:2016ynf} using the effective LRSM. The content of this model consists of the usual SM states, the $W^{\prime \pm}$ and $Z^\prime$ gauge bosons, which are aligned with their mass eigenstates, and three heavy Majorana neutrinos $N_i$, aligned with the RH chiral states. In the LRSM, the $W^\prime$ chiral coupling to quarks is given by
\begin{equation}
	\mathcal{L}_{W^\prime q q^\prime}=\frac{-\kappa^q_R g}{\sqrt{2}} \sum_{i,j=u,d,...}\bar{u_i} V^{^\prime}_{ij} W^+_{R \mu} \gamma^\mu P_R d_j+\text{H.c.},
\end{equation}
where $\kappa^q_R$ is a coefficient associated with the coupling of the $Z^\prime$ boson to fermions, $u_i(d_j)$ is an up-(down-)type quark of flavor $i(j)$; $P_{R(L)}=\frac{1}{2}(1 \pm \gamma^5)$ denotes the RH(LH) chiral projection operator, and $V^{\prime}_{ij}$ is the RH Cabibbo-Kobayashi-Maskawa (CKM) matrix, which is related to the SM CKM matrix.
\begin{table}[!t]
\begin{center}
\begin{tabular}{ c | c | c  c  c  c  c  c  c  c }
\hline \hline
Gauge Group	&	Charge	     	&	$u_L$	& $d_L$	& $\nu_L$ & $e_L$ 	& $u_R$	& $d_R$	& $N_R$ & $e_R$ \\
\hline
$SU(2)_L$	&	$T_{L}^{3,f}$	&	$+\frac{1}{2}$ & $-\frac{1}{2}$ & $+\frac{1}{2}$ & $-\frac{1}{2}$ & 0 & 0 & 0 & 0 \\

$SU(2)_R$	&	$T_{R}^{3,f}$	&	0 & 0 & 0 & 0 & $+\frac{1}{2}$ & $-\frac{1}{2}$ & $+\frac{1}{2}$ & $-\frac{1}{2}$ \\

$U(1)_{\rm EM}$	&	$Q^f$		&	$+\frac{2}{3}$ & $-\frac{1}{3}$ & $0$ & $-1$ & $+\frac{2}{3}$ & $-\frac{1}{3}$ & $0$ & $-1$ 
\tabularnewline\hline
\hline
\end{tabular}
\end{center}
\caption{SU$(2)_L$, SU$(2)_R$, and U$(1)_{\rm EM}$ quantum number assignments for chiral fermions $f$ in LRSM.}
\label{tb:qNumbers}
\end{table}

For leptons, the $W^\prime$ coupling and leptonic mixing is parametrized by
\begin{equation}
	\mathcal{L}_{W^\prime l \nu/N}=\frac{-\kappa^l_R g}{\sqrt{2}}\sum_{l=e,\mu,\tau}\left[\sum^3_{m=1}\bar{\nu}^c_m X_{lm}+\sum^3_{m^\prime=1} \bar{N}_{m^\prime} Y_{l m^\prime} \right] W^{\prime +}_{\mu} \gamma^\mu P_R l^- + \text{H.c.} .
\end{equation}
The matrix $Y_{l m^\prime}(X_{lm})$ quantifies the mixing between the heavy (light) neutrino mass eigenstate
$N_{m^\prime}(\nu_m)$ and the RH chiral state with corresponding lepton flavor $l$. The mixing scale as
\begin{equation}
	|Y_{l m^\prime}|^2 \sim \mathcal{O}(1) \quad \text{and} \quad |X_{l m}|^2 \sim 1-|Y_{l m^\prime}|^2 \sim \mathcal{O}(m_{\nu_m}/m_{N_{m^\prime}})  .
\end{equation}
As in the quark sector, $\kappa^l_R \in \mathbb{R}$ independently normalizes the strength of $W^\prime$ coupling to leptons.\par
After LR symmetry breaking, the $W^3_R$ and $X_{(B-L)}$ gauge states mix and give rise to the massive $Z^\prime$ and massless (hypercharge) $B$ bosons. Subsequently, all fermions with $(B-L)$ charges, including $\nu_L$ and $N_R$, couple to $Z^\prime$. For chiral fermion $f$, we parametrize the $Z^\prime$ neutral currents by
\begin{equation}\label{CouplingZpff}
	\mathcal{L}_{Z^\prime f f} = \frac{-\kappa_R^f g}{\sqrt{1-\left(1/\kappa^f_R\right)^2 \tan^2 \theta_W}}\sum_{f=u,e,...}\bar{f}Z_{\mu}^\prime \gamma^\mu \left(g^{Z^\prime,f}_L P_L + g_R^{Z^\prime,f} P_R\right)f.
\end{equation}
where $\kappa^f_R$ is the same $\kappa^{q,f}_R$ as for $W_R$.  In terms of electric and isospin charges, the chiral coefficients are
\begin{equation}\label{gL}
	g^{Z^\prime,f}_L=\left(T^{3,f}_L - Q^f \right)\frac{1}{\kappa^{f 2}_R} \tan^2 \theta_W,
\end{equation}
\begin{equation}\label{gR}
	g^{Z^\prime,f}_R=T^{3,f}_R -\frac{1}{\kappa^{f 2}_R}\tan^2 \theta_W Q^f .
\end{equation}
$SU(2)_L$, $SU(2)_R$, and $U(1)_{EM}$ quantum number assignments for $f$ are summarized in Table \ref{tb:qNumbers}. 
We provide the neutral current expression for $Z^\prime$ as 
\begin{equation}\label{currentEQ}
\begin{split}
	J^{\mu}_{Z^\prime} &= \sum_{f=u,e,...}\bar{f}Z_{\mu}^\prime \gamma^\mu \left(g^{Z^\prime,f}_L P_L + g_R^{Z^\prime,f} P_R\right)f\\
&=\frac{1}{2}\sum_{f=u,e,...}\bar{f}Z_{\mu}^\prime \gamma^\mu \left[c_V^f-c_A^f \gamma^5 \right]f,
\end{split}
\end{equation}
where $c_{V(A)}^f=g_L^{Z^\prime,f} +(-) g_R^{Z^\prime,f}$ are the corresponding vector and axial couplings. For generic $\kappa^f_R$ normalizations, the LO $Z^\prime$ partial decay width is 
\begin{equation}
\begin{split}\label{widthEq}
 \Gamma(Z^\prime \to f\bar{f}) &=  N_c^f \frac{\kappa_{Z^\prime}^{f 2} g^2 M_{Z^\prime} \sqrt{1-4r_f^{Z^\prime}}}{48\pi\left[1 - (1/\kappa_{R}^f)^2 \tan^2\theta_W\right]}\\
 &\times
 \left[(g_{L}^{Z^\prime,f}+g_{R}^{Z^\prime,f})^2(1+2r_f^{Z^\prime})+(g_{L}^{Z^\prime,f}-g_{R}^{Z^\prime,f})^2(1-4r_f^{Z^\prime})\right],
\end{split}
\end{equation}
where {$r_i^{V}=m_i^2/M^2_{V}$. The total cross section for $Z^\prime$ is denoted as $\Gamma_{Z^\prime}=\sum_f\Gamma(Z^\prime\ \to f \bar{f})$. We present contour plots illustrating  the decay width of $Z^\prime$ projected onto $\kappa^f_R -M_{Z^{\prime}}$ plane, as shown in Figure \ref{WidthContour}.
\begin{figure}[h] 
\centering
\includegraphics[width=9cm,height=6cm]
{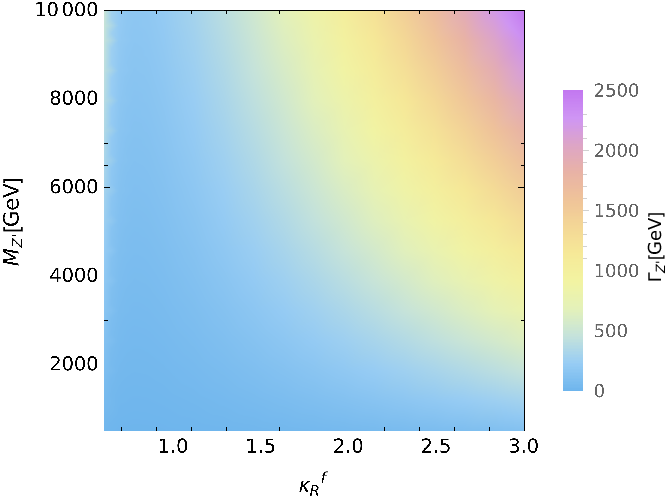}
\caption{
Contour plots of the total decay width of $Z^\prime$ corresponding to different values of $\kappa^f_R$ and $M_{Z^\prime}$.}
\label{WidthContour}
\end{figure}
Based on Equation \eqref{CouplingZpff}, the  coupling must satisfy $\lvert \kappa^f_R \rvert>0.55$. Notably, when $\kappa^f_R$ approaches 0.55, it leads to a significantly larger and unphysical decay width for $Z^\prime$. 
It is observed that for higher values of $\kappa^f_R$ (e.g., $\kappa^f_R > 2$), relatively large decay widths are also incurred for high-mass $Z^\prime$. 

In order to visually comprehend the impact of $\kappa^f_R$ variation on the coupling strengths $g^f$ of $Z^\prime$ with different fermions, Figure \ref{glim} illustrates how the coupling strengths between $Z/Z^\prime$ and different fermions vary with $\kappa^f_R$. The lighter color lines represent the coupling strengths between $Z$ boson and the corresponding particles in the SM. The absolute value of coupling strength lies between 0.25 to 0.74 in the SM. The coupling strengths of $Z^\prime$ are plotted with the dark color lines. The value of $\kappa^f_R \gtrsim 0.55$ is a limit imposed by $|\tan \theta_W/\kappa^f_R|<1$. According to Equation \eqref{gR},  the  coupling strengths of $Z^\prime$ and quarks satisfy  $g_{L_{u,c,t}}=g_{L_{d,s,b}}$.
\begin{figure}[h] 
\centering
\includegraphics[width=10cm,height=6cm]
{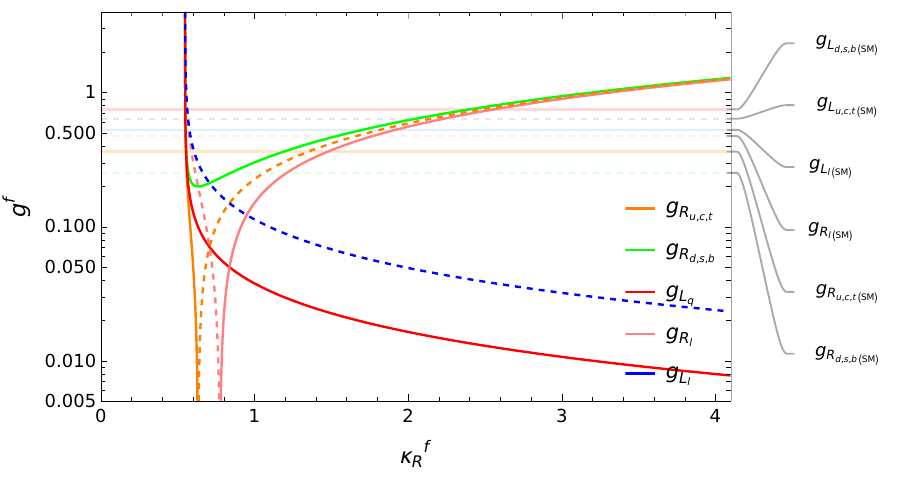}
\caption{ The coupling strengths between $Z^\prime/Z$ and different fermions vary with $\kappa^f_R$. The  lighter color lines stand for the couplings of Standard Model $Z$ boson with fermions and the dark color ones stand for the couplings of $Z^\prime$ boson. Especially,  the solid lines represent the coupling is positive, while the colored dashed lines represent the absolute values of the corresponding coupling strengths in the negative values region.
}
\label{glim}
\end{figure}
In Figure \ref{glim}, 
the intersection point of the orange dashed line and the red solid line indicates a partial cancellation of the coupling terms between left-handed and right-handed particles in the $Z^\prime \to q \bar{q}(q=u,c,t)$ process. This point corresponds to the minimum value of the cross section for this process, which is 1.58 pb. Similarly, when the pink solid line and the blue dashed line intersect, the $\mu^+ \mu^- \to Z^\prime \to l^+ l^-$ process attains its minimum cross section of 0.132 pb at $M_{Z^\prime}$ = 6 TeV, with a total decay width of 120 GeV and a collision energy of 6 TeV.
\begin{table}[tbh]
\begin{center}
\setlength{\abovecaptionskip}{6pt}
\setlength{\belowcaptionskip}{0pt}
\begin{tabular}{ c c c c c c c}
\hline
\hline
Collaboration & $\sqrt{s}$[TeV] & $\mathcal{L}$[fb$^{-1}$] & $l_{final}$ & $M_{N_{l}}$[TeV] &Upper limit of $M_{Z^\prime}$[TeV] & Ref. \\
\hline
ATLAS & $8.00$ & $20.30$ & $e^{\pm}$ & $0.34$ & $2.01$ & \cite{ATLAS:2015gtp}  \\
CMS & $13.00$ & $138$ & $e^{\pm}$ & $0.31$ & $3.83$ & \cite{CMS:2023ooo} \\
ATLAS & $8.00$ & $20.30$ & $\mu^{\pm}$ & $0.27$ & $2.20$ & \cite{ATLAS:2015gtp}  \\
CMS & $13.00$ & $138$ & $\mu^{\pm}$ & $0.22$ & $4.42$ & \cite{CMS:2023ooo} \\
ATLAS & $8.00$ & $20.30$ & $e^{\pm}$ or $\mu^{\pm}$ & $0.29$ & $2.23$ & \cite{ATLAS:2015gtp}  \\
\hline
\hline
\end{tabular}
\caption{The upper limits for $Z^\prime$ based on the $Z^\prime \to N_l N_l \to l^\pm l^\pm W^{\prime \mp *} W^{\prime \mp *}$ process at the LHC, where $M_{N_l}$ is the neutrino mass and $l_{final}$ is the type of final state lepton.}
\label{LHCLimit}
\end{center}
\end{table}}
The presence of both heavy gauge boson and right-handed neutrino is a  unique feature of LRSM.   A typical decay mode of  $Z^\prime \to N_R N_R \to l^\pm l^\pm W^{\prime \mp *} W^{\prime \mp *} $ has been elaborately studied at the LHC recently. The mass constraints of   $Z^\prime$ have been listed in  Table \ref{LHCLimit} with the final states including electrons or muons. The observed upper limit of $Z^\prime$ mass  reaches up to 4.42 TeV in the LRSM~\cite{CMS:2023ooo}. Additionally, the mass limitation of $Z^\prime$ has been surveyed in various other models, as summarized in detail in reference~\cite{Lu:2023jlr}.  Due to the limitation of the current collision energy and integrated luminosity at the LHC, the observed mass range for $Z^\prime$ is limited to less than 5 TeV. 
\section{$Z^\prime$ Boson Production and Decay at Future Muon Collider}
\begin{figure}
\centering
\includegraphics[width=8cm,height=6cm]{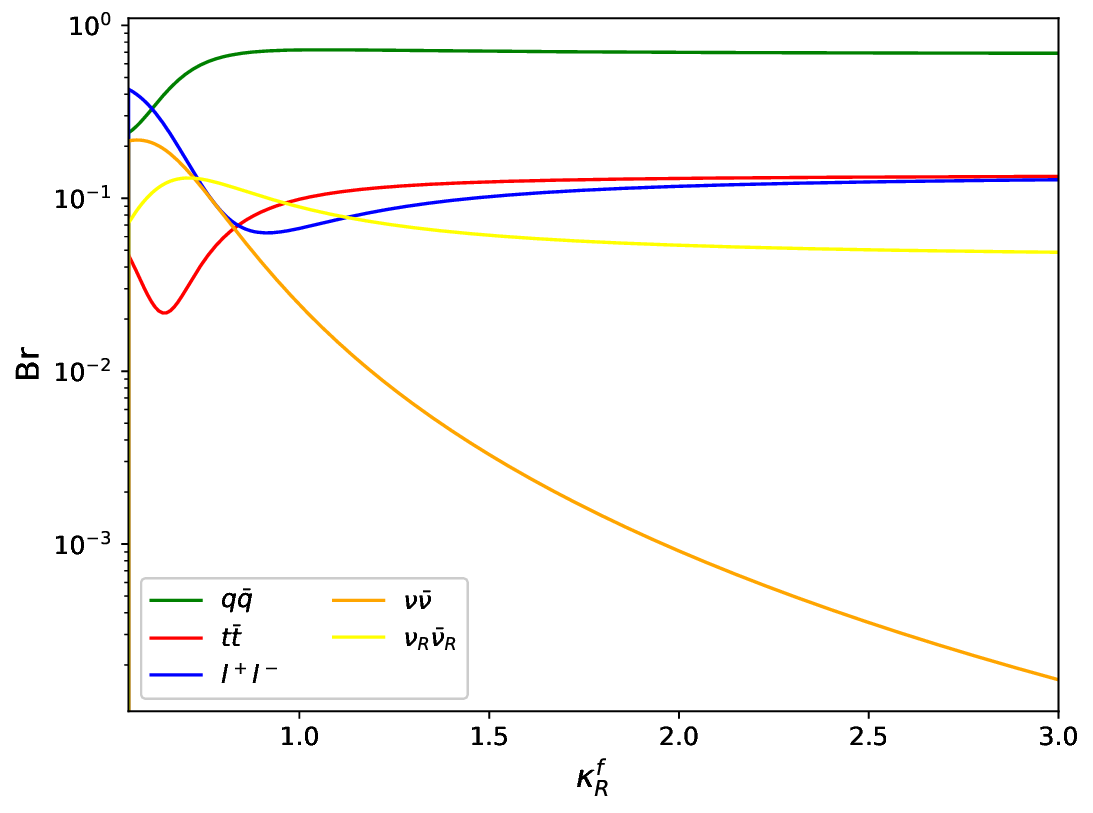}
\caption{The decay branching ratios of various decay channels of $Z^\prime$ as functions of $\kappa^f_R$.}
\label{BR}
\end{figure}

In LRSM, the new neutral gauge boson $Z^\prime$ can couple with the SM particles as well as the right-handed neutrinos. The  right-handed neutrino can be a good candidate for dark matter in the studies of dark sector. However, in this paper we mainly focus on the interactions of $Z^\prime$  coupling to fermions in the SM. The typical mass is set as $m_{Z^\prime}= 6$ TeV, $m_{\nu_R}=173.3$ GeV for the lightest right-handed neutrino, the other right-handed neutrinos are much heavier than 6 TeV and can be neglected in this study.
We present the $Z^\prime$ decay branching ratios as functions of $\kappa^f_R$, as shown in Figure \ref{BR}.
The branching ratios of different decay channels exhibit significant fluctuations for $\kappa^f_R<1$, while they stabilize when $\kappa^f_R>1.5$. Among them, $\mu^+ \mu^- \to q \bar{q}$, $\mu^+ \mu^- \to t \bar{t}$ and $\mu^+ \mu^- \to l^+ l^-$ are the dominant decay channels.  For example, with $\kappa^f_R = 0.6$, the decay branching ratios for the $Z^\prime \to q \bar{q}$, $Z^\prime \to t \bar{t}$, and $Z^\prime \to l^+ l^-$ decay channels are $41\%$, $23\%$, and $4.5\%$, respectively. However, when the value of $\kappa^f_R$ is 2, the decay branching ratios for these three decay channels become 69\%, 13\%, and 11\%.
These primary decay channels will  be elaborately studied in our work.

\begin{figure}
  \centering

  \begin{minipage}{0.49\textwidth}
    \centering
    \includegraphics[width=7.5cm]{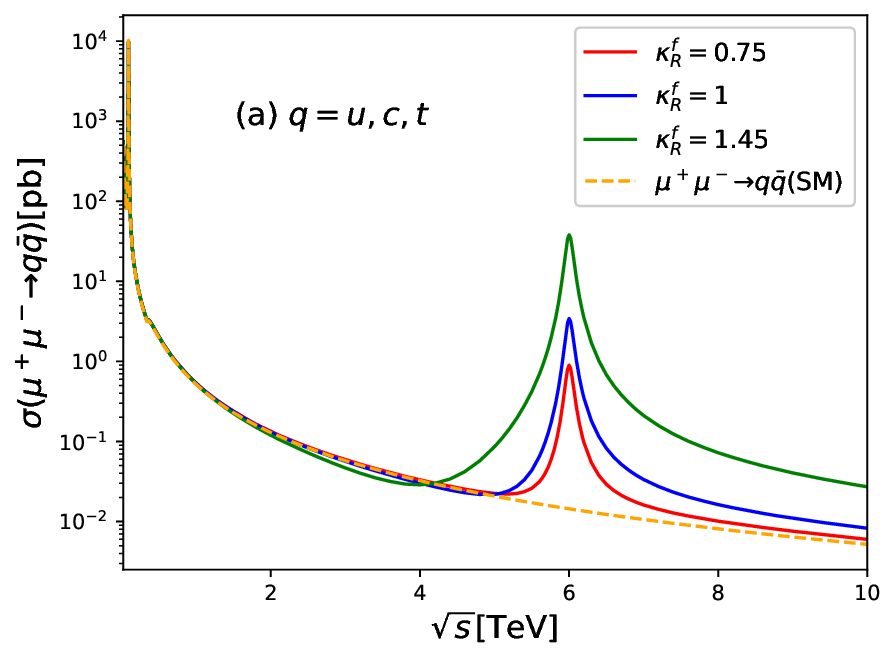}
  \end{minipage}\hfill 
  \begin{minipage}{0.49\textwidth}
    \centering
    \includegraphics[width=7.5cm]{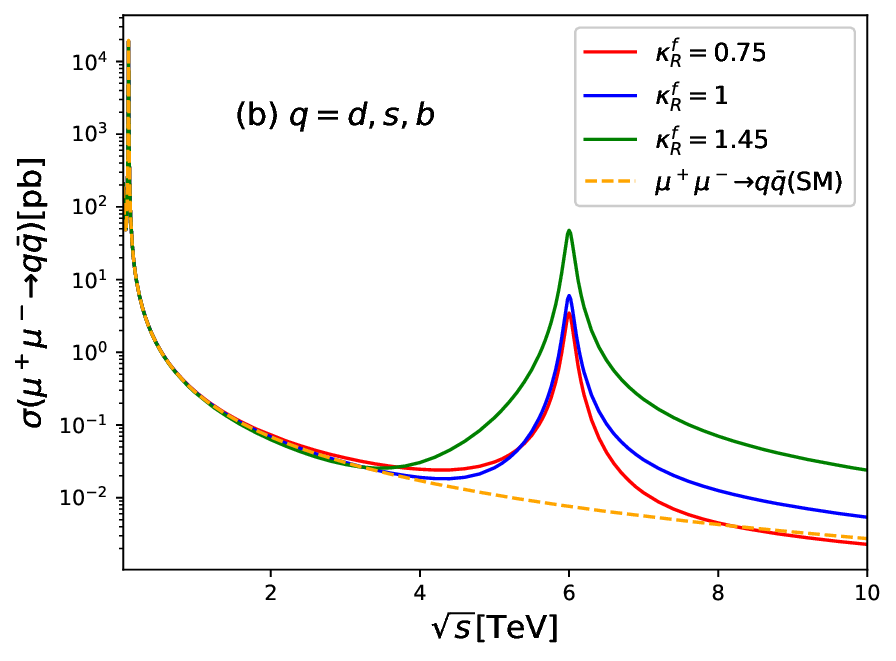}
  \end{minipage}
  \begin{minipage}{0.49\textwidth}
    \centering
    \includegraphics[width=7.5cm]{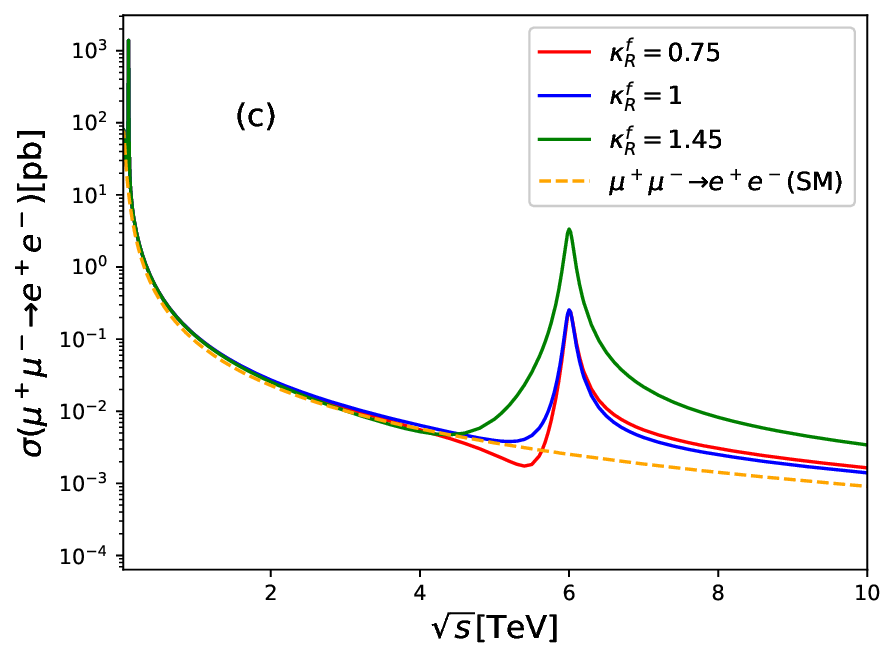}
  \end{minipage}\hfill 
  \begin{minipage}{0.49\textwidth}
    \centering
    \includegraphics[width=7.5cm]{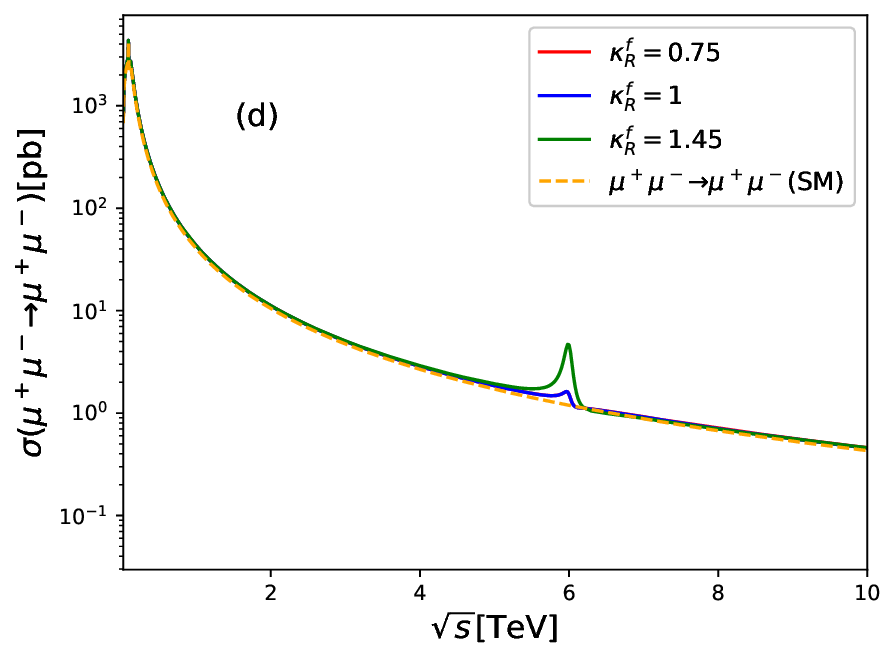}
  \end{minipage}

  \caption{The variation of the scattering cross section for the process $\mu^+ \mu^- \to q \bar{q}$ (a), (b), $\mu^+ \mu^- \to e^+ e^-$ (c) and $\mu^+ \mu^- \to \mu^+ \mu^-$ (d) with collision energy, where $M_{Z^\prime}$=6 TeV and $\Gamma_{Z^\prime}$=120 GeV.}
  \label{mumuCro}
\end{figure}

The cross section distributions with  $Z^\prime$ induced processes are shown in Figure \ref{mumuCro} with different collision energy.  The mass of $Z^\prime$ is set to 6 TeV as an example and the decay  width is fixed to $2\%$ of its mass. The processes $ \mu^+\mu^- \to q \bar{q} $ (a) with up-type quarks, (b) with down-type quarks, $\mu^+\mu^- \to e^+e^- $ (c), and $ \mu^+\mu^- \to \mu^+\mu^- $ (d) are investigated with $\kappa^f_R$ is set to 0.75, 1, and 1.45. We categorize the  $q = u,c,t$ and $q = d,s,b$ into separated process for the study of the interaction of $Z^\prime q \bar{q}$. The resonance peak is obviously with $\sqrt{s}=m_{Z^\prime}$ in each plot. The magnitude of the resonance peak depends on the contribution of the $Z^\prime$ to the total scattering cross-section in the vicinity of the resonance state. The cross section will be increased with the coupling strength $\kappa^f_R$ enlarged.  For the process $ \mu^+\mu^- \to Z^\prime \to q \bar{q} $, as evident in Figure \ref{glim}, it is straightforward to observe that the primary influence on the processes with $q = u,c,t$ and $q = d,s,b$ comes from the coupling of $Z^\prime$ with right-handed quarks. This leads to subtle differences in Figures \ref{mumuCro} (a) and (b). For Figures \ref{mumuCro} (a) and (b), before reaching the resonance energy, i.e., within a partial interval where the collision energy is about between 2 and 4 TeV, the total cross-section for $ \mu^+\mu^- \to \gamma^*/Z/Z^\prime \to q \bar{q} $ is smaller than the total cross-section for $ \mu^+\mu^- \to \gamma^*/Z \to q \bar{q} $. This phenomenon is attributed to the negative contribution from the interference of $Z-Z^\prime$ and $Z^\prime-\gamma^*$. The interference terms are denoted as
\begin{equation}\label{Mmplitude}
  \mathcal{M}^2_{Z^\prime}+2(\mathcal{M}_{Z^\prime} \mathcal{M}_{\gamma^*}+\mathcal{M}_{Z^\prime} \mathcal{M}_{Z})<0,
\end{equation}
where $\mathcal{M}_{Z}$, $\mathcal{M}_{Z^\prime}$, and $\mathcal{M}_{\gamma^*}$ are the amplitudes of the process $ \mu^+\mu^- \to q \bar{q} $ with $Z$, $Z^\prime$, and $\gamma^*$ as the propagator,  respectively.
 With $\kappa^f_R>0.75$, in the range where the collision energy is greater than the resonance mass, especially greater than 8 TeV, the cross section $\sigma(\mu^+ \mu^- \to Z/Z^\prime/\gamma^* \to q \bar{q})$ is less than $\sigma(\mu^+ \mu^- \to Z/\gamma^* \to q \bar{q})$ in Figure \ref{mumuCro} (b). When the C.M.S energy is less than the resonance mass, the interference term of Equation \eqref{Mmplitude} is greater than 0, however the $Z^\prime$ propagator term is negative in this energy region. These two factors provide a negative contribution to the scattering cross section of this decay channel.

Figures \ref{mumuCro} (c) and (d) show the cross section distribution of  processes with double electrons and double muons as final states, respectively. Although both are leptonic final states, the scattering cross sections for these two processes exhibit significant differences with varying collision energy. In Figure \ref{mumuCro} (c), the process with double electrons as the final state shows a scattering cross section distribution that is greatly influenced by the interference effects of $Z^\prime-Z$ and $Z^\prime-\gamma^*$ around the resonance peak. For instance, with $\kappa^f_R = 0.75$ and $ \sqrt{s}=5.4$ TeV, a pronounced minimum of $1.74 \times 10^{-3}$ pb is observed in the scattering cross section. This phenomenon is attributed to the interference effects between $Z^\prime$ and $Z$, and between $Z^\prime$ and $\gamma^*$, similar to the double quark process. In comparison to the double-electron process, the resonance peak produced by the double-muon process in Figure \ref{mumuCro} (d) is relatively smaller but still sufficient to detect a resonant signal at the collider. Another noteworthy observation is that, in the double-muon process, regions outside the resonance state are minimally influenced by $Z^\prime$. This is attributed to the fact that the double-muon process includes contributions not only from the s-channel Feynman diagrams but also from the t-channel Feynman diagrams. The existence of t-channel Feynman diagrams mitigates the interference of $Z^\prime$ with the s-channel diagrams of $Z$ boson and $\gamma^*$, rendering the signal of $Z^\prime$ less prominent in regions that deviate from the resonance  peak energy by more than 2 TeV.

The scattering cross section of the $\mu^+ \mu^- \to q\bar{q} $ process depends on the coupling strength of the $ Z^\prime \mu^+\mu^- $ vertex as well as the coupling strength of  the $ Z^\prime q\bar{q} $ vertex. In Figure \ref{mumuCro}, we assume that the values of $ \kappa_R^q $ and $ \kappa_R^l $ are equal. However, in many theoretical scenarios, $ \kappa_R^q $ and $ \kappa_R^l $ may have distinct values. Therefore, it is crucial to investigate the combined impact of  $ \kappa_R^q $ and $ \kappa_R^l $ in the scattering cross section of the $ \mu^+ \mu^- \to q\bar{q} $ process. 
\begin{figure}
  \centering

  \begin{minipage}{0.49\textwidth}
    \centering
    \includegraphics[width=7.5cm]{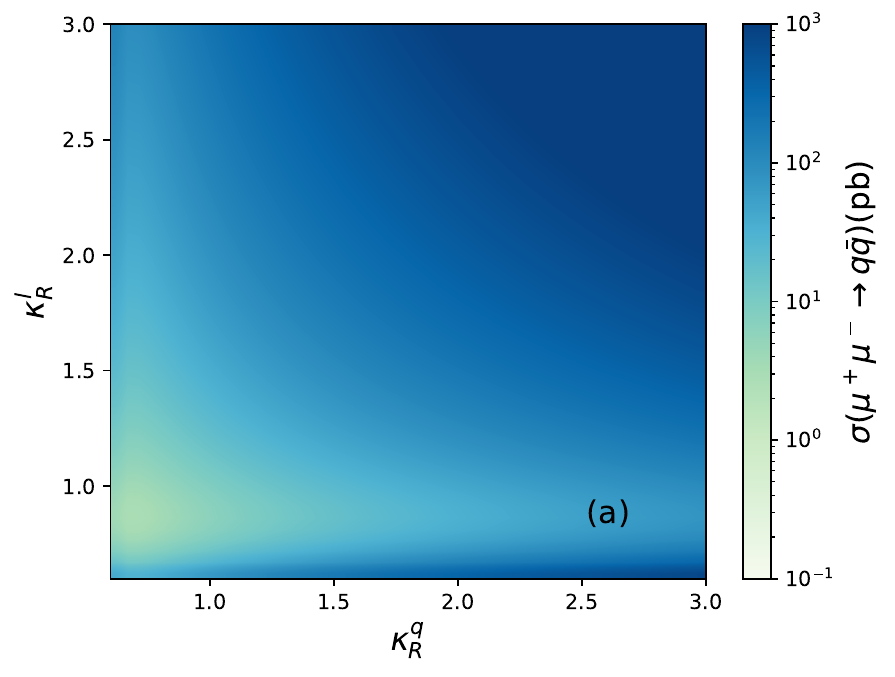}
  \end{minipage}\hfill 
  \begin{minipage}{0.49\textwidth}
    \centering
    \includegraphics[width=7.5cm]{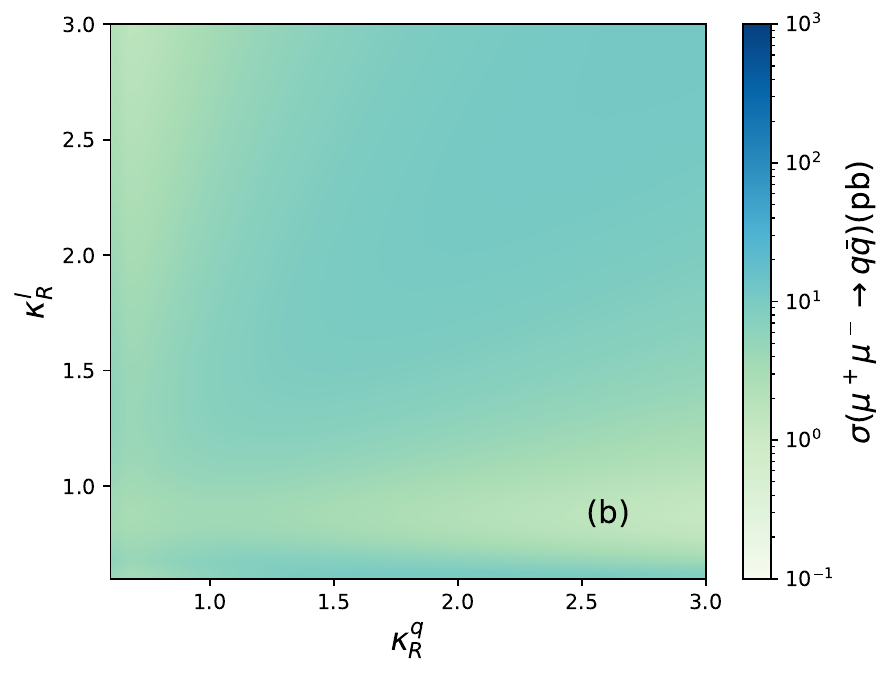}
  \end{minipage}

  \caption{ Contour plots of the scattering cross section for the process $\mu^+\mu^- \to q\bar{q}(q=u,c,d,s,b)$ with different values of $\kappa_R^q$ and $\kappa_R^l$. 
  The collision energy is set to 6 TeV as well as the mass of $Z^\prime$, and the decay width of $Z^\prime$ is set at 120 GeV (a) and  calculated by the formula of Equation~\eqref{widthEq} (b).}
  \label{diffCoupingCroContour}
\end{figure}
In Figure \ref{diffCoupingCroContour}, we present contour plots illustrating the scattering cross section distribution of the $ \mu^+ \mu^- \to q\bar{q} $ process corresponding to various values of $ \kappa_R^q $ and $ \kappa_R^l $. To better illustrate the impact of $\kappa_R^q$ and $\kappa_R^l$ on the scattering cross section for this process, we present Figure \ref{diffCoupingCroContour} with the collision energy equal to the resonance mass. Considering the significant influence from the decay width of $Z^\prime$ on this process, we present the results with a fixed width in Figure \ref{diffCoupingCroContour} (a) and calculated the total decay width using Equation \eqref{widthEq} in Figure \ref{diffCoupingCroContour} (b). Due to the resonance effects,  the cross section is larger than $ 1 $ pb  in most regions of $0.6<\kappa_R^f<3.0$.  However, variations in the cross section are also evident with changes in $\kappa_R^q$ and $\kappa_R^l$. 

In Figure \ref{diffCoupingCroContour} (a) with a fixed decay width, the cross section gradually increases with the increase of $\kappa_R^q $ and $ \kappa_R^l$.   The minimum value is 2.8 pb with $\kappa_R^q = 0.65, \kappa_R^l = 0.84$. This distribution is precisely consistent with the distribution depicted in Figure \ref{glim}, where the lines of $g_{L_q}$ intersect with $-g_{R_{u,c,t}}$, and $g_{R_l}$ intersect with $-g_{L_l}$. In the vicinity of these two values, the coupling strengths for $Z^\prime q \bar{q}$ and $Z^\prime l^+ l^-$ both reach their minimum values.  In Figure \ref{diffCoupingCroContour} (b), the cross section distributions versus couplings are plotted with a calculated decay width according to Equation \eqref{widthEq}.
The cross section is large in the region where both $\kappa_R^q$ and $\kappa_R^l$ are large. While in the region of left-top corner and right-bottom corner in the plot, the cross section is smaller than other regions because of the large decay width of $Z^\prime$ suppressing the cross section increasing with $\kappa_R^f$. In this scenario, the minimum value occurs at $\kappa_R^q = 3$ and $\kappa_R^l = 0.844$, with a value of 0.368 pb. 

\begin{figure}
  \centering

  \begin{minipage}{0.49\textwidth}
    \centering
    \includegraphics[width=7.5cm]{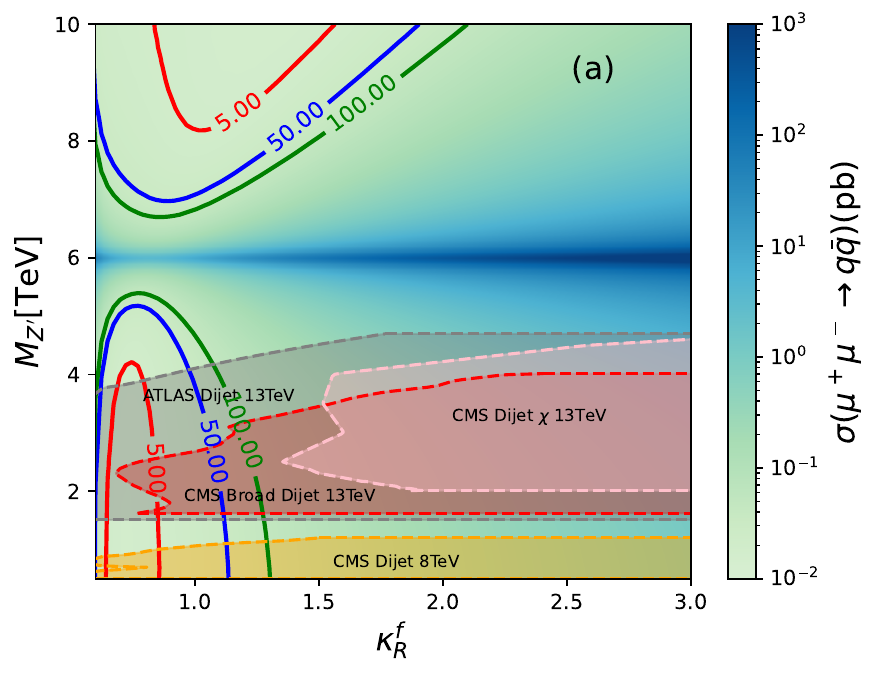}
  \end{minipage}\hfill 
  \begin{minipage}{0.49\textwidth}
    \centering
    \includegraphics[width=7.5cm]{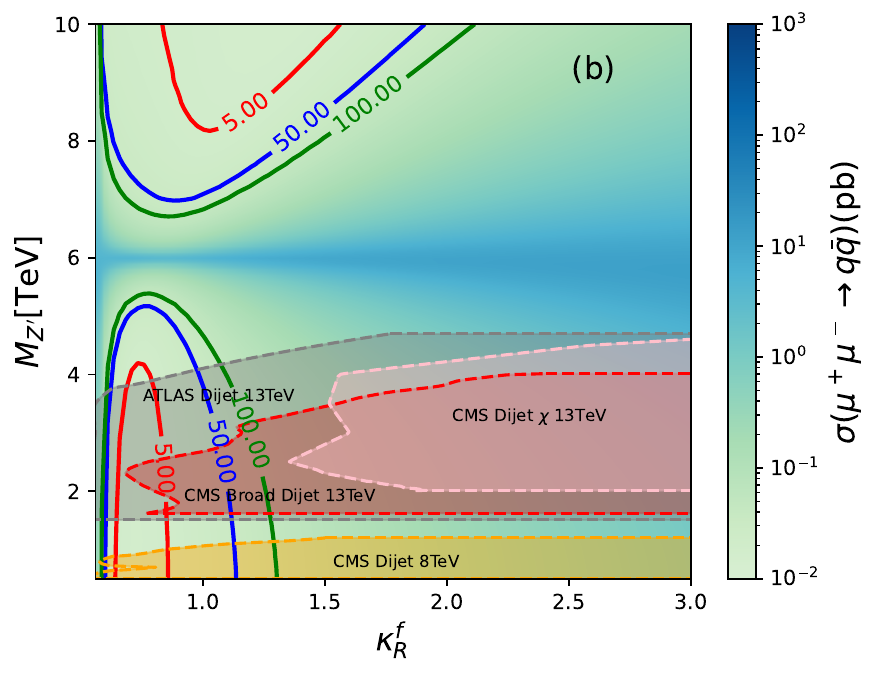}
  \end{minipage}

  \caption{The contour plots show the cross section for the process $\mu^+\mu^- \to q\bar{q}(q=u,c,d,s,b)$ with different values of $\kappa_R^f$ and $M_{Z^\prime}$. The solid lines in red, blue, and green represent significance of 5, 50 and 100, respectively. The coloured regions of pink\cite{CMS:2018ucw}, gray\cite{CMS:2018mgb}, red\cite{ATLAS:2019fgd}, and orange\cite{CMS:2018kcg}  are the experimental excluded regions from the LHC with different collision energy.  The total collision energy and the mass of $Z^\prime$ are both set at 6 TeV and the decay width of $Z^\prime$ is $2\%$ of $M_{Z^\prime}$ (a) and calculated by the formula of Equation~\eqref{widthEq} (b).}
  \label{mumuToqqCroCon}
\end{figure}

In Figure \ref{mumuToqqCroCon}, we present the impact of $Z^\prime$ mass and $\kappa^f_R$ on the scattering cross section of the $\mu^+ \mu^- \to q \bar{q}$ process under different decay width scenarios: (a) the decay width of $Z^\prime$ is $2\%$ of $M_{Z^\prime}$,  (b) calculating the decay width using Equation \eqref{widthEq}.  
Additionally, we map the  constraints from current LHC results on this process in the plot, with the experimental limits imposed on the $Z^\prime \to q \bar{q}$ coupling strength $g^\prime_q$. The experimental constraints are concentrated within the $Z^\prime$ mass range of $10 - 3000$ GeV. The colored regions stand for the experimental results with different collision energies. 
 The variations in the  decay width of $Z^\prime$ have an impact on the values of the scattering cross section shown in Figure 6 (a) and (b), but have little effect on the overall distribution of the scattering cross section. This is due to the assumption that $ \kappa_R^q = \kappa_R^l $ in this process, resulting in a symmetric distribution of cross section represented in Figure \ref{diffCoupingCroContour} (a) and (b) along the diagonal line from the bottom-left corner to the top-right corner. As the collision energy reaches 6 TeV, the  cross section rapidly increases near $Z^\prime$ = 6 TeV due to the resonant effects. This rapid increase enhances the excess of the new physics signal out of the standard model process. The significance (defined as $N_{Z^\prime}/\sqrt{N_{SM}}$) of the scattering cross section for the $\mu^+ \mu^- \to q \bar{q}$ process is provided in the presence of $Z^\prime$ boson.  One can find that most regions in the plane of $\kappa_R^q - m_Z^{\prime}$ can be detected except the left-bottom and left-top corner regions with significance less than five at the future muon colliders. Compared with the current experimental results at the LHC, it shows the significant advantage of detecting  massive $Z^\prime$ boson at the future muon collider. 
For a 6 TeV muon collider, the exclusion mass region of $Z^\prime$ reaches at least 8 TeV with a large $ \kappa_R^q $.  For colliders with higher collision energies, the exclusion range for the $Z^\prime$ mass will be further expanded. For a 10 TeV collider, the exclusion range for $Z^\prime$ mass could potentially reach 20 TeV or even higher.\par

\begin{figure}
  \centering

  \begin{minipage}{0.49\textwidth}
    \centering
    \includegraphics[width=7.5cm]{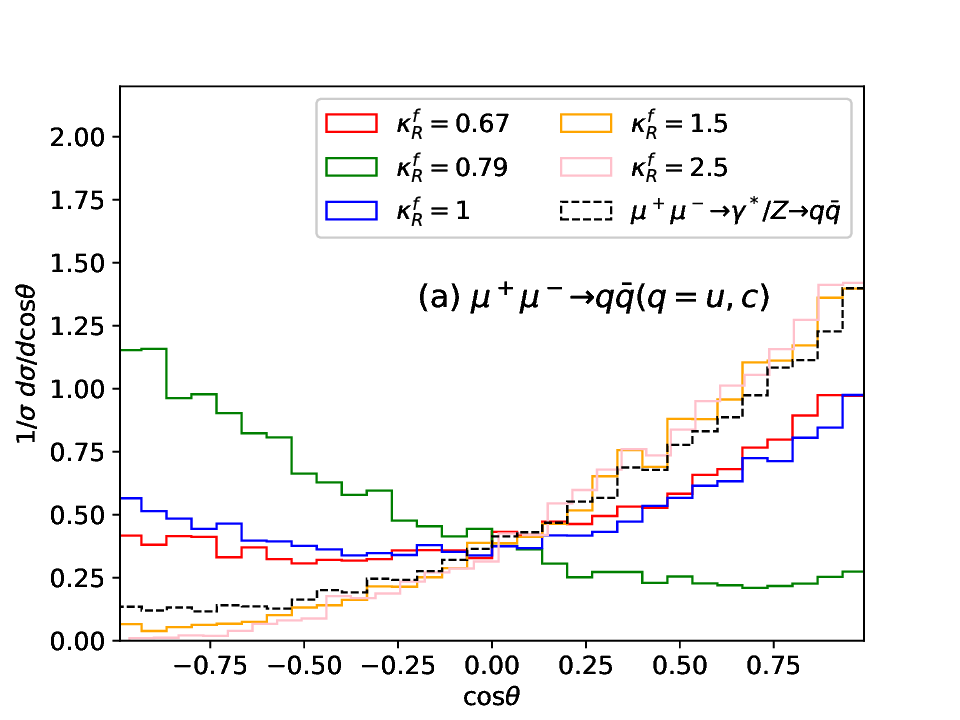}
  \end{minipage}\hfill 
  \begin{minipage}{0.49\textwidth}
    \centering
    \includegraphics[width=7.5cm]{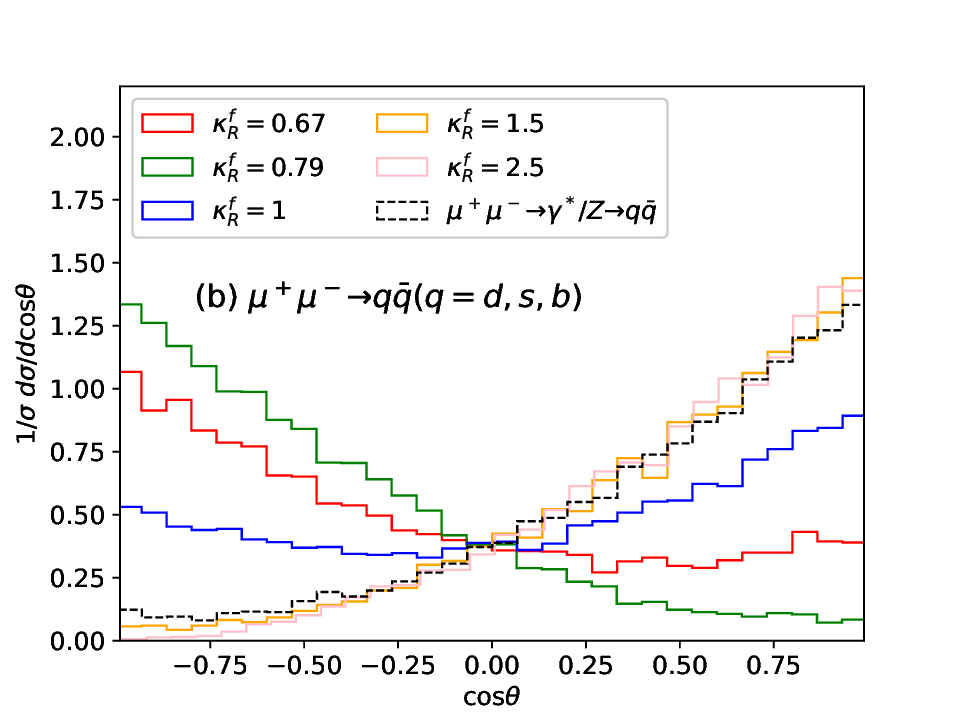}
  \end{minipage}
  \begin{minipage}{0.49\textwidth}
    \centering
    \includegraphics[width=7.5cm]{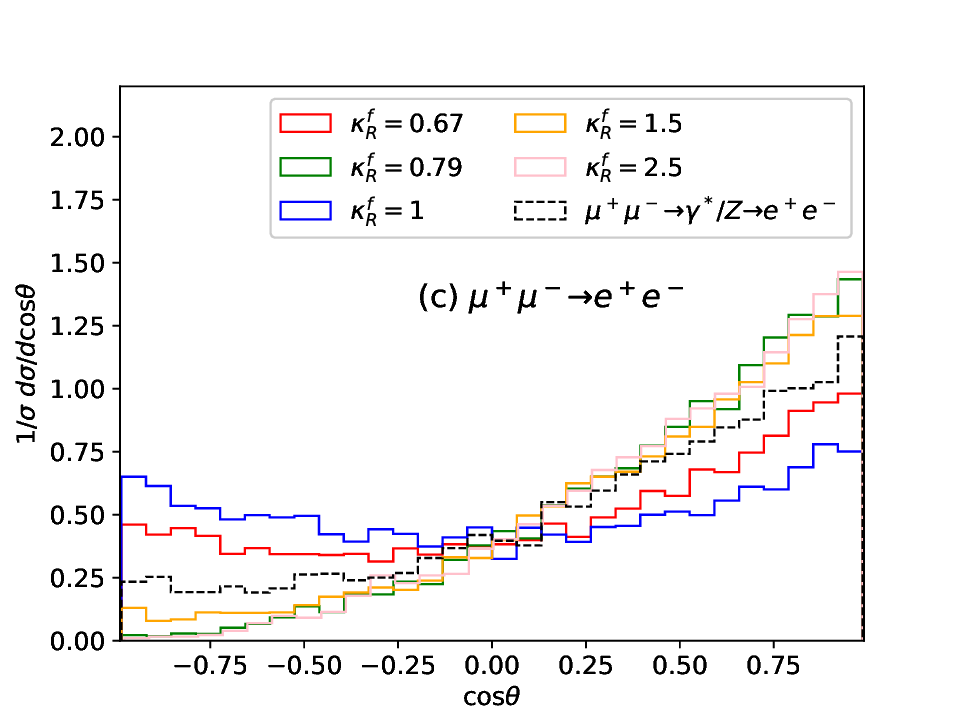}
  \end{minipage}\hfill 
  \begin{minipage}{0.49\textwidth}
    \centering
    \includegraphics[width=7.5cm]{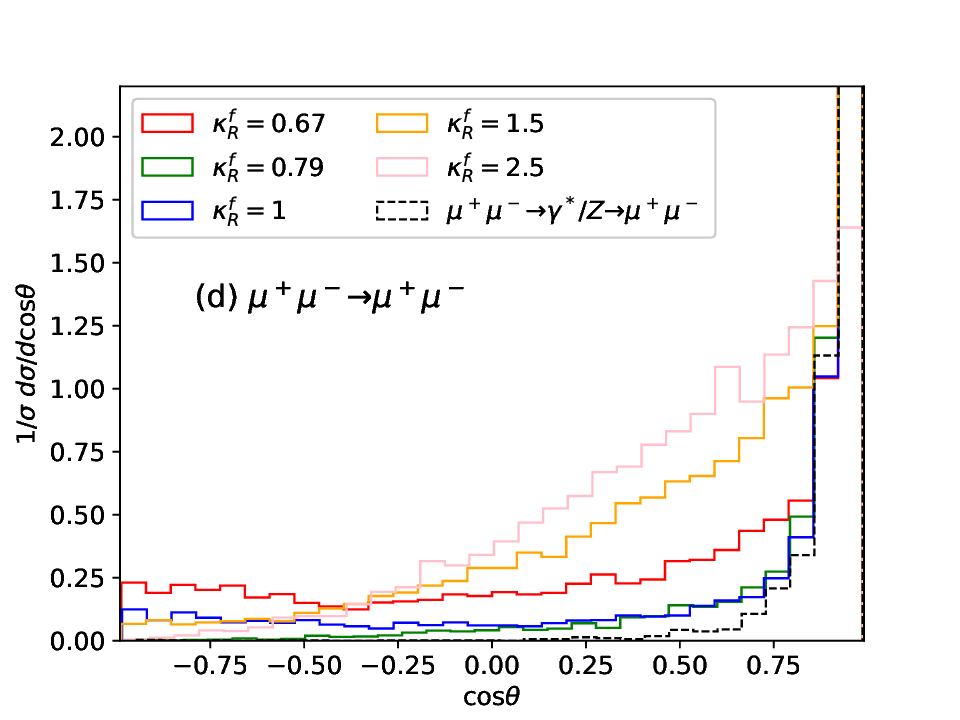}
  \end{minipage}

  \caption{ Angular distributions of final-state particles in the processes $\mu^+ \mu^- \to q\bar{q} $ (a), (b), and $ \mu^+ \mu^- \to l^+ l^- $ (c), (d) corresponding to key values of $\kappa^f_R$.}
  \label{mumuDis}
\end{figure}

In addition to the scattering cross section, we have also conducted an in-depth study of the angular distribution of final-state particles in various decay channels of $Z^\prime$ at a future muon collider. The angle between initial and final-state particles is defined as 
\begin{equation}\label{cosTheta}
    \cos\theta=\frac{\bm{p_{f}^*} \cdot \bm{p_{i}}}{|\bm{p_{f}^*}|\cdot |\bm{p_{i}} |},
\end{equation}
where $\bm{p_{f}^*} $ and $\bm{p_{i}}$ are the three momentum of the final and initial particle, respectively. In Figure \ref{mumuDis}, we present angular distributions of final-state particles for the processes $ \mu^+ \mu^- \to q\bar{q} $ (a), (b), and $ \mu^+ \mu^- \to l^+ l^- $ (c), (d) corresponding to different values of $ \kappa^{f}_R $. 
Comparing the angular distributions of processes including $Z^\prime$ with the corresponding processes in the Standard Model, significant changes occur in the angular distributions of final-state particles in the range of $ \kappa^{f}_R < 1.5 $. 

We also define the forward-backward asymmetry of the final-state particle distributions using the formula
\begin{equation}\label{AFB}
    A_{FB}=\frac{\sigma(\cos\theta \geq 0)-\sigma( \cos \theta <0)}{\sigma(\cos\theta \geq 0)+\sigma( \cos \theta <0)}.
\end{equation}
The corresponding asymmetry values for the angular distributions in Figure \ref{mumuDis} are listed in Table \ref{FBATable}. In Figure \ref{mumuDis} and Table \ref{FBATable}, it is evident that the final-state particle angular distributions of $\mu^+ \mu^- \to q \bar{q}$ and $\mu^+ \mu^- \to l^+ l-$ are sensitive to changes in the $\kappa^f_R$ values when $\kappa^f_R$ is less than 1.5. If there exists a $Z^\prime$ with $\kappa^f_R$ less than 1.5, it could be easily observed at future muon colliders through anomalous final-state particle angular distributions and asymmetry in the angular distributions. This sensitivity provides a valuable observable for detecting signals of TeV-scale $Z^\prime$ particles. For instance, in Figure \ref{mumuDis} (a), (b), the values of $\kappa^f_R$ at 0.67, 0.79, and 1 result in a reduction in asymmetry, even reaching negative values. The same trend of change is shown in \ref{mumuDis} (c), (d), where distinct differences in the final-state lepton angular distributions are also observed. This phenomenon greatly helps in the understanding and detection of right-handed neutral gauge bosons.

\begin{table}[!t]
\begin{center}
\setlength{\abovecaptionskip}{6pt}
\setlength{\belowcaptionskip}{0pt}
\begin{tabular}{ c c c c c}
\hline
\hline
$\kappa^f_R$ & $A_{FB}[q \bar{q}(u,c)]$ & $A_{FB}[q \bar{q}(d,s,b)]$ &  $A_{FB}[e^+ e^-]$ & $A_{FB}[\mu^+ \mu^-]$\\
\hline
0.67 & $0.290$ & $-0.314$ & $0.262$ & $0.649$ \\
0.79 & $-0.472$ & $-0.667$ & $0.726$ & $0.969$ \\
1 & $0.186$ & $0.196$ & $0.052$ & $0.848$ \\
1.5 & $0.700$ & $0.700$ & $0.66$ & $0.733$ \\
2.5 & $0.751$ & $0.743$ & $0.738$ & $0.746$ \\
SM & $0.611$ & $0.648$ & $0.488$ & $0.998$ \\
\hline
\hline
\end{tabular}
\caption{Forward-backward asymmetry for the process $\mu^+ \mu^- \rightarrow q \bar{q}$ and $\mu^+ \mu^- \rightarrow l^+ l^-$ with $m_{Z^\prime}=6$ TeV. The SM values are listed in the last line.}
\label{FBATable}
\end{center}
\end{table}
\begin{figure}
  \centering
  \begin{minipage}{0.49\textwidth}
    \centering
    \includegraphics[width=7.5cm]{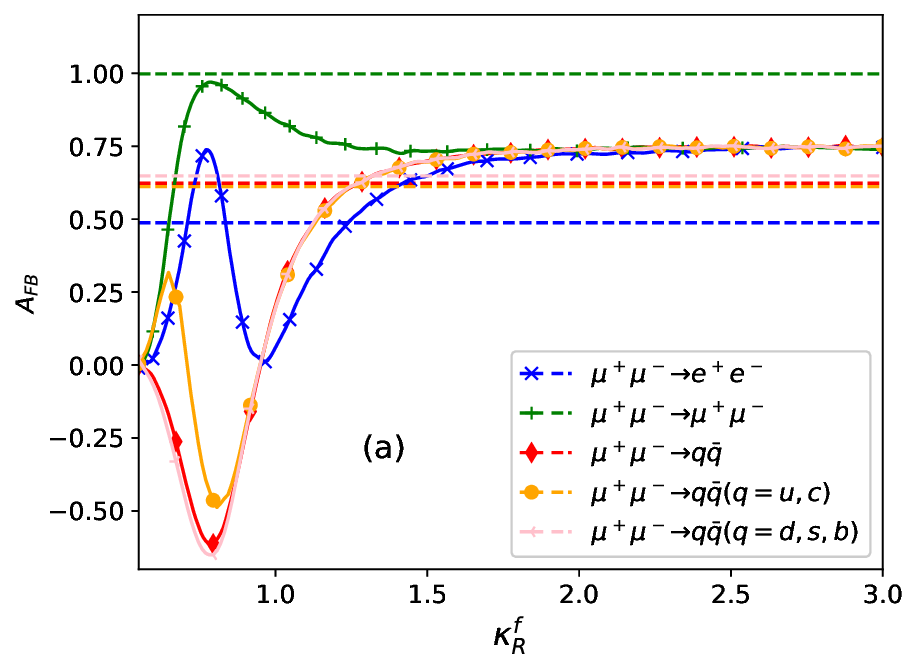}
  \end{minipage}\hfill 
  \begin{minipage}{0.49\textwidth}
    \centering
    \includegraphics[width=7.5cm]{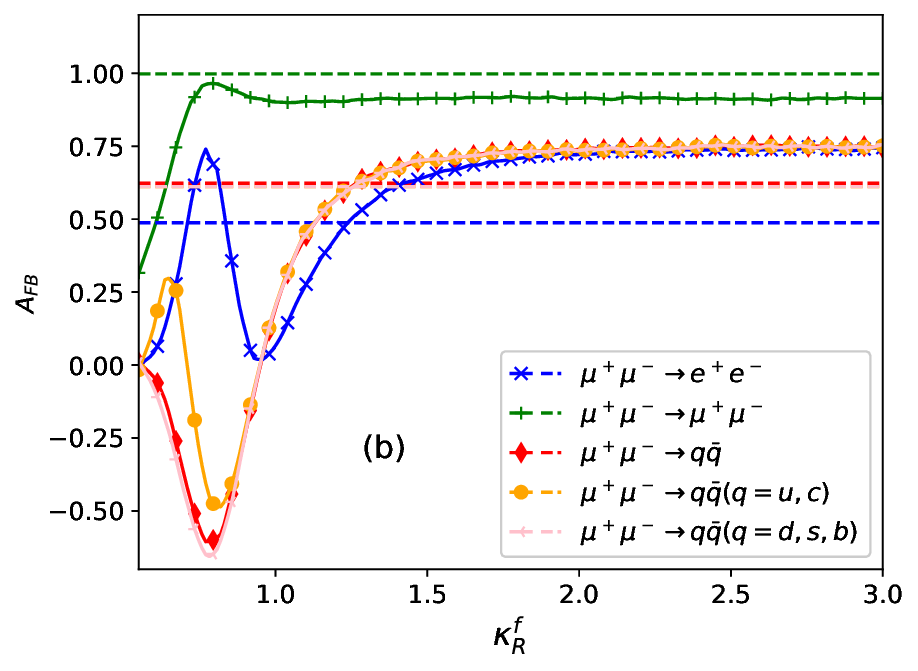}
  \end{minipage}
  \caption{The dependence of $A_{FB}$ on $\kappa^f_R$ for a muon collider with a center-of-mass energy of 6 TeV and $Z^\prime$ mass of 6 TeV. The decay width of $Z^\prime$ is $2\%$ of $M_{Z^\prime}$ (a) and calculated by the formula of Equation~\eqref{widthEq} (b).}
  \label{AFB2D}
\end{figure}
To investigate the impact of $\kappa^f_R$ variations on the final-state particle angular distributions, we provide the forward-backward asymmetry of the angular distributions for different decay channels by varying the $\kappa^f_R$ values. The asymmetry of each decay channel with respect to the change in $\kappa^f_R$ is illustrated in Figure \ref{AFB2D}. 
In Figure \ref{AFB2D} (a), the total decay width of $Z^\prime$ is set to 120 GeV, while in Figure \ref{AFB2D} (b), the decay width is calculated by Equation \eqref{widthEq}. The lines in Figure \ref{AFB2D} indicate that, for small values of $\kappa^f_R$ ($\kappa^f_R < 1.5$), there is a significant variation in the asymmetry of the final-state angular distribution with changing $\kappa^f_R$. 
This means if a right-handed $Z^\prime$ boson with a small $\kappa^f_R$  value exists, one could extract the value of $\kappa^f_R$ by measuring the asymmetry in the angular distribution of final-state particles at future muon colliders. This observation is consistent with the conclusion presented in Figure \ref{mumuDis}. In the region where $\kappa^f_R$ is less than 1.5, there are several turning points, and one of the turning points for the asymmetry change trend in all decay channels occurs around $\kappa^f_R$ = 0.79. This is because the value of $g_{R_l}$ approaches 0 in the vicinity, and subsequently, as the $\kappa^f_R$ value increases, the value of $g_{R_l}$ changes from negative to positive. Since $g_{R_l}$ plays a role not only in the dilepton process but also affects the diquark process at future muon colliders, this turning point is significant. Apart from this point, there is another transition point for $\mu^+ \mu^- \to q \bar{q}(q=u,c)$ and $\mu^+ \mu^- \to e^+ e^-$ processes, corresponding to $\kappa^f_R$ values around 0.75 and 1, respectively. This is because with $\kappa^f_R \approx 0.75$, the sum of $g_{L_q}$ and $g_{R_{u,c,t}}$ is close to zero, reducing the contributions from $Z^\prime$ to this process. A similar phenomenon occurs for $g_{L_l}$ and $g_{R_l}$ around $\kappa^f_R \sim  1$. However, the double muon process has different distributions because of the additional three Feynman diagrams from the $t-$channel, which reduce the impact of $\kappa^f_R$ on the asymmetry in this process compared to the double electron final state. Furthermore, for $\kappa^f_R$ values greater than 1.5, the asymmetry in the dilepton decay channel still exhibits notable deviations from the Standard Model. As a whole,  the asymmetry tends to a constant with a large $\kappa^f_R$ for all decay channels. This phenomenon is attributed to the stabilization of the coupling strength between $Z^\prime$ and fermions as $\kappa^f_R$ increases, along with the gradual increase in contributions from Feynman diagrams where $Z^\prime$ serves as the propagator.
\begin{figure}
  \centering

  \begin{minipage}{0.49\textwidth}
    \centering
    \includegraphics[width=7.5cm]{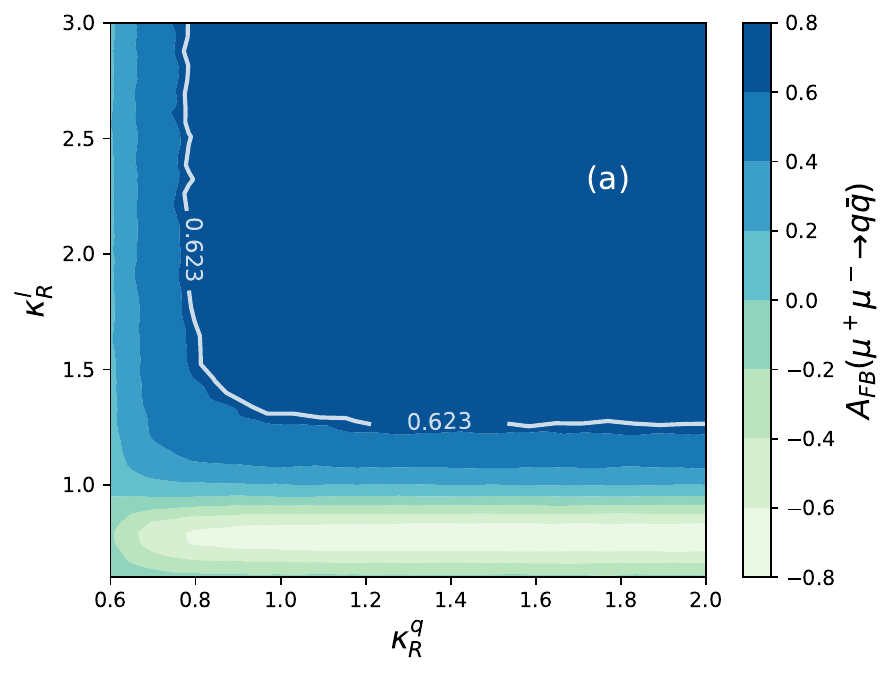}
  \end{minipage}\hfill 
  \begin{minipage}{0.49\textwidth}
    \centering
    \includegraphics[width=7.5cm]{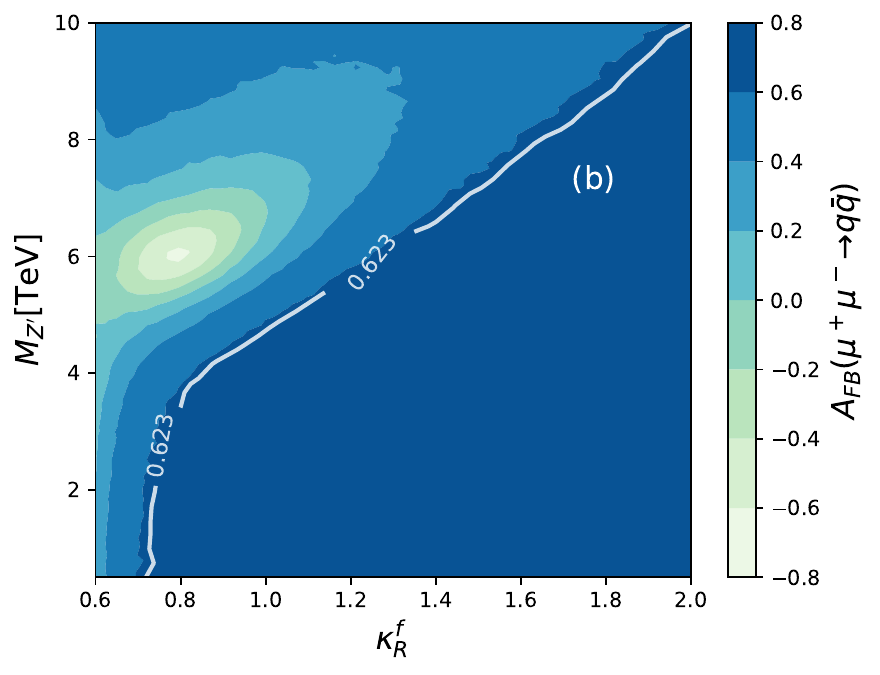}
  \end{minipage}

  \caption{The contour plots of the asymmetry of the final-state quark angular distribution vary with $ \kappa_R^{q} $ and $ \kappa_R^{l} $ (a) as well as $ \kappa^f_R $ and $ M_{Z^\prime} $ (b)  in the $ \mu^{+}\mu^{-} \to q\bar{q} $ process. Here, $M_{Z^\prime}=6$ TeV for (a) and $\sqrt{s} = 6 $ TeV,  $ \Gamma_{Z^\prime} $ calculated by the formula of Equation~\eqref{widthEq}. The white line represents the asymmetry of the final-state quark angular distribution in the standard model for this process, with a value of 0.623.
}
  \label{AFBqq3D}
\end{figure}

The final state angular distribution of $ \mu^+ \mu^- \to q \bar{q} $ process depends on the parameters of $ \kappa_R^q $, $ \kappa_R^l $, and $ M_{Z^\prime}$. In figure \ref{AFBqq3D} (a) we present the asymmetry contour plots with different values of $ \kappa_R^q $ and $ \kappa_R^l $. The plot indicates that when $ \kappa_R^q $ is small $ (\kappa_R^q < 1 )$, its variation significantly influences the asymmetry of the final-state particle angular distribution in this process. As $ \kappa_R^q $ increases, its impact on the asymmetry decreases. The main impact on the asymmetry of the final-state particle angular distribution in this process comes from $ \kappa_R^l $, where $ g_{R_l} $ and $ -g_{L_l} $ intersect near $ \kappa_R^l = 1 $. After the intersection, $ g_{R_l}$ plays a predominant role with $\left |{g_{R_l}}\right |\textgreater \left | g_{L_l}\right |$.This corresponds to a significant change in the asymmetry of the final-state particle angular distribution in $ \mu^+ \mu^- \to q \bar{q} $. 
 Figure \ref{AFBqq3D} (b) shows contour plots of the asymmetry as a function of $ \kappa_R^f $ and $ M_{Z^\prime} $, where $ \kappa_R^f = \kappa_R^l = \kappa_R^q $.  The negative asymmetry observed in the interval of small $ \kappa_R^f $, which corresponds to the left-bottom region in  Figure \ref{AFBqq3D} (a). When $ M_{Z^\prime} $ is 6 TeV, the process will be affected by the  resonance effect, and the contribution of $ Z^\prime $ reaches its maximum. This effectively explains the  minimum value of the asymmetry around $ \kappa_R^l \sim 0.8$ and $ M_{Z'} \sim 6$  TeV.

\begin{figure}
  \centering

  \begin{minipage}{0.49\textwidth}
    \centering
    \includegraphics[width=7.5cm]{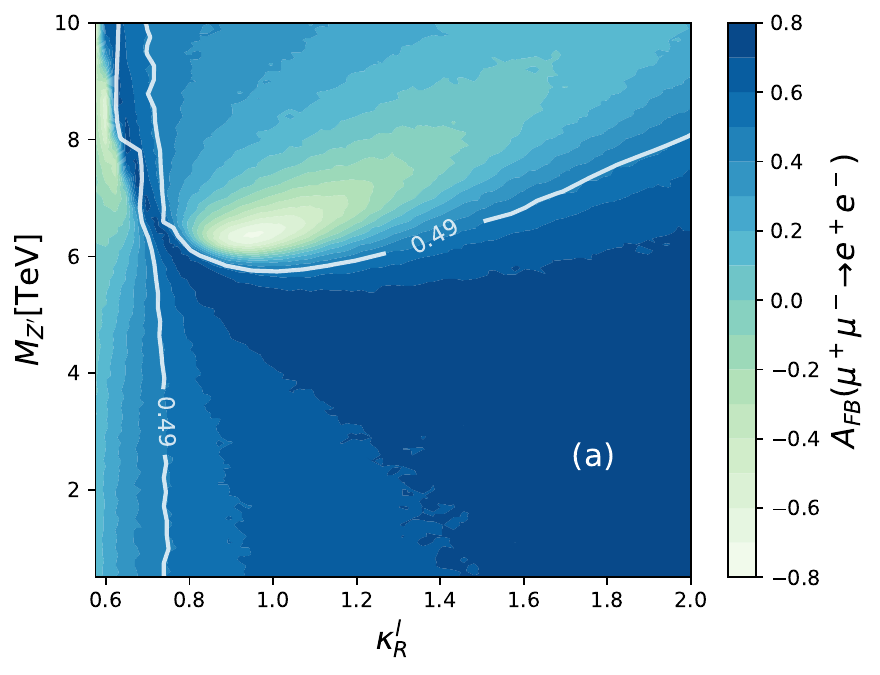}
  \end{minipage}\hfill 
  \begin{minipage}{0.49\textwidth}
    \centering
    \includegraphics[width=7.5cm]{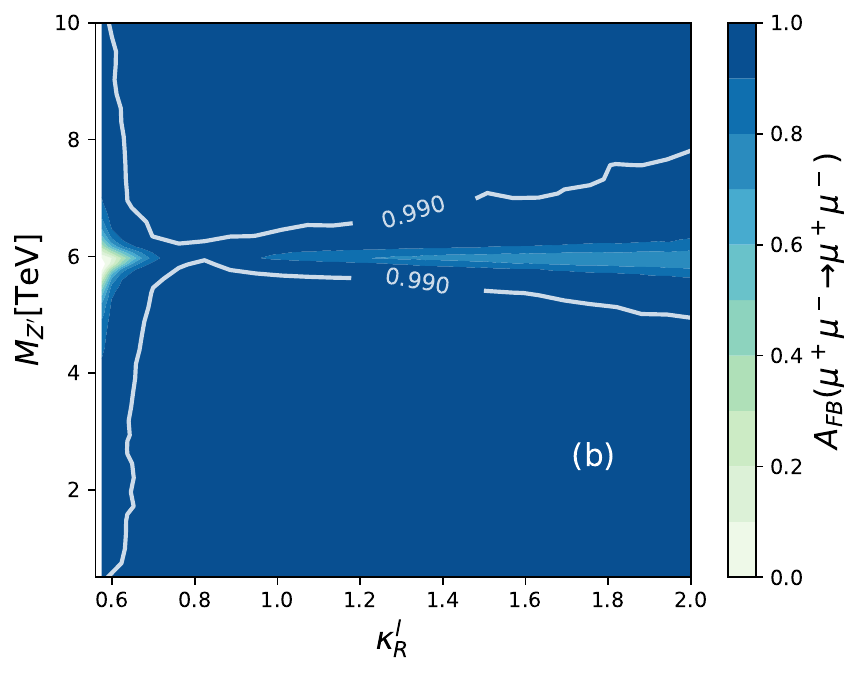}
  \end{minipage}

  \caption{The contour plots of the asymmetry of the final-state quark angular distribution vary with $ \kappa^l_R $ and $ M_{Z^\prime} $ in the process of $ \mu^+ \mu^- \to e^+ e^-  $ (a) and $ \mu^+ \mu^- \to \mu^+ \mu^- $ (b). The collision energy is set 6 TeV and the total width of $Z^\prime$ is set to $2\%$ of the $Z^\prime$ mass. The solid white lines represent the asymmetry of the final-state lepton angular distribution in the Standard Model for the corresponding processes. The asymmetry is 0.49 in (a) and 0.99 in (b).}
  \label{AFBll3D}
\end{figure}

 The angular distributions of final-state leptons are different from the distributions  of final-state quarks.
In Figure \ref{AFBll3D} (a), the asymmetry of the final-state lepton angular distribution in the process $\mu^+\mu^-\to e^+e^-$ is depicted as a function of $\kappa^l_R$ and $M_{Z^\prime}$. The distribution is consistent with  Figure \ref{AFB2D}, where the impact of $\kappa^l_R$ on the asymmetry is evident in this process. Specifically, when $\kappa^l_R$ is around 0.8, the asymmetry rapidly increases to an extremum, then decreases rapidly with further increments in $\kappa^l_R$. A prominent peak forms around $\kappa^l_R = 0.8$, followed by a pronounced minimum near $\kappa^l_R = 1$. The significant influence of $Z^\prime$ on this process reaches its maximum effect when $\kappa^l_R = 1$ and gradually diminishes. It is noteworthy that in the region where $M_{Z'}$ is below 6 TeV, the asymmetry rapidly increases to values close to those of the Standard Model. In contrast, in the region where $M_{Z^\prime}$ exceeds 6 TeV, the growth of the asymmetry is slower, resulting in elliptical contour lines. This observation is consistent with the behavior of this process in the interval where $\kappa^l_R$ is greater than 1, characterized by a rapid increase in $Z^\prime$'s contribution when the collision energy is below the resonance energy and approaches it. 
In Figure \ref{AFBll3D} (b), the asymmetry of the final-state lepton angular distribution in the process $\mu^+\mu^- \to \mu^+\mu^-$ is presented. This distribution shows a great discrepancy with the process of electron final-state. Except in the vicinity of the resonance peak production with $\sqrt{s}\sim m_{Z^\prime}$, the asymmetry in other regions tends toward the values of the Standard Model. This is attributed to the introduction of additional $t-$channel Feynman diagrams in this process, where the inclusion of $t-$channel Feynman diagrams leads to a substantial increase in the asymmetry of the final-state lepton angular distribution. This effect is further manifested in a significant reduction in the impact of $Z^\prime$ on the asymmetry of this process. Thus, investigating $Z^\prime$ in the $\mu^+\mu^- \to \mu^+\mu^-$ process is more challenging than in the $\mu^+\mu^- \to q\bar{q}$ and $\mu^+\mu^- \to e^+e^-$ processes.
\section{Summary}
In this paper, we investigate the phenomenology of the neutral gauge boson $Z^\prime$ and its coupling to standard model fermions in the context of the Left-Right Symmetric Model (LRSM) at future muon colliders.   The research of the $Z^\prime$ boson at the LHC has provided the result of  the mass region with a few TeV. In this work we show  the superiority of  $Z^\prime$ search at the future muon collider with mass region exceed the current hadron colliders . 

The key  parameters in the production and decay of $Z^\prime$  boson include the coupling parameter $\kappa_R^f$ , the total width $\Gamma_{Z^\prime}$ and the mass of $Z^\prime$. 
Typical values for $\kappa_R^f$ are chosen as 0.75, 1, and 1.45, representing different coupling strength regimes within the model's allowed range. The typical mass of $Z^\prime$ is set to  6 TeV, therefore, the resonance effect can be investigated with collision energy  of $\sqrt{s}=6$ TeV. We provide the total decay width of $Z^\prime$ through two methods: $2\%$ of $Z^\prime$ mass or calculation with the theoretical Formula \eqref{widthEq}. The primary decay channels of $Z^\prime$ are $Z^\prime \to q\bar{q}$ and $Z^\prime \to l^+ l^-$. Compared to the neutral gauge bosons in the standard model, the extra neutral gauge boson $Z'$ exhibits intriguing phenomena. The production of $Z^\prime$ shows significant enhancement in both special quark and lepton couplings. The scattering cross sections for the $\mu^+\mu^- \to q\bar{q}$ (where $q = u, c, t$), $q\bar{q}$ (where $q = d, s, b$), $e^+e^-$, and $\mu^+\mu^-$ processes at the resonance peak are 38.1 pb, 47.61 pb, 3.36 pb, and 4.63 pb, respectively. Each decay channel exhibits unique characteristics, and the muon collider provides a larger exclusion space for the parameter range of $Z^\prime$ mass. For instance, the detection capability for the diquark process reaches up to 10 TeV or even higher at a 6 TeV muon collider. Furthermore, for future muon colliders with collision energies of 10 TeV or even exceeding 10 TeV, the prospect of detecting $Z^\prime$ particles with masses of 20 TeV or higher is anticipated.

The  final-state particle angular distributions are sensitive observables in various decay channels for the investigation of $Z^\prime$ interactions. The study reveals that within the range of $\kappa^f_R $ from 0.55 to 1.5, changes in $\kappa^f_R $ have a significant impact on the angular distributions of final-state particles in different decay channels. As shown in Figure \ref{AFB2D}, these changes are determined by the specific coupling terms of the right-handed neutral gauge boson $Z^\prime$ in the LRSM, as given by equations \eqref{CouplingZpff}, \eqref{gL}, and \eqref{gR}. These distinctive structures result in a non-linear relationship between $ g^\prime$ and $ \kappa^f_R $, and the dependence of different coupling constants on various gauge group quantum numbers leads to more complex variations in the couplings, as depicted in Figure \ref{glim}. The contributions of $Z^\prime$ to different production processes vary due to the differences in collision energy and $Z^\prime$ mass. We defined a  forward-backward asymmetry  from the angular distribution of final-state particles to show the property of $Z^\prime$ boson. Specifically, near the resonance state, when $ \kappa^{f_R} $ is in a smaller range, the asymmetry is significantly reduced, and in some cases, it exhibits completely opposite symmetry compared to the SM process. For example, in the $\mu^+\mu^- \to q\bar{q} $ process, the forward-backward asymmetry  is 0.611 (0.648) for  $q = u, c$ ( $q = d, s, b$) in the SM but 0.29 (-0.314) including $Z^\prime$ contributions with $ \kappa^f_R =0.67$. This is a consequence of the coupling of the right-handed $Z^\prime$ with fermions. Similar conclusions apply to the $ \mu^+\mu^- \to e^+e^- $ process, with a broader range of reduced asymmetry. However, the impact on the asymmetry  is limited by $Z^\prime$  in  the $ \mu^+\mu^- \to \mu^+\mu^- $ process due to the contributions from the $t-$channel. The results imply that in future muon colliders, distinguishing the presence of a large mass neutral gauge boson can be achieved by examining the forward-backward asymmetry of the scattering angles of final-state quarks or leptons in different energy ranges. The study of extra neutral gauge bosons at a muon collider not only provides new methods for discovering new physics but also reveals prospects for future muon colliders.

\section*{Ackonwledgement}
This work was supported by the Natural Science Foundation of Shandong Province under grant Nos.~ZR2022MA065,  ZR2023MA013, ZR2021QA040, and National Natural Science Foundation of China under Grant No. 12105162.

\bibliographystyle{apsrev4-2}
\bibliography{Reference}
\end{document}